\documentclass[twocolumn,showpacs,preprintnumbers,amsmath,amssymb,prb,superscriptaddress]{revtex4}
\usepackage{graphicx}
\usepackage{dcolumn}
\usepackage{bm}
\usepackage{epstopdf}

\begin{document}
\title{Extended Hubbard model on a C$_{20}$ molecule}
\author{Fei Lin}
 \affiliation{Department of Physics, University of Illinois at Urbana-Champaign, Urbana, Illinois 61801, USA}
\author{Erik S. S{\o}rensen}
 \affiliation{Department of Physics and Astronomy, McMaster University, Hamilton, Ontario,
 Canada L8S 4M1}
\author{Catherine Kallin} 
 \affiliation{Department of Physics and Astronomy, McMaster University, Hamilton, Ontario,
 Canada L8S 4M1}
\author{A. John Berlinsky}
 \affiliation{Department of Physics and Astronomy, McMaster University, Hamilton, Ontario,
 Canada L8S 4M1}
 \date{\today}

\begin{abstract}
The electronic correlations on a C$_{20}$ molecule, as described by 
an extended Hubbard Hamiltonian with a nearest neighbor Coulomb interaction of strength $V$, 
are studied using quantum Monte Carlo and exact diagonalization methods. 
For electron doped C$_{20}$, it is known that pair-binding arising from a purely electronic 
mechanism is absent within the standard Hubbard model ($V=0$). Here we show that this is 
also the case for hole doping for $0<U/t\leq 3$ and that, for both electron and hole doping,
the effect of a non-zero $V$ is to work against pair-binding. We also study the magnetic 
properties of the neutral molecule, and find transitions between spin singlet and triplet 
ground states for either fixed $U$ or $V$ values. In addition, spin, charge and pairing 
correlation functions on C$_{20}$ are computed. The spin-spin and charge-charge correlations 
are very short-range, although a weak enhancement in the pairing correlation is observed for 
a distance equal to the molecular diameter.
\end{abstract}


\maketitle

\section{Introduction}
Shortly after the discovery of superconductivity in C$_{60}$, it
was suggested by Chakravarty, Kivelson and Gelfand
\cite{kivelson91a, kivelson91b, kivelson01} that an electronic mechanism, in
which pairs of electrons preferentially reside on a single
molecule rather than on neighboring molecules, might provide the
pairing mechanism for superconductivity. Using second order perturbation theory
they found evidence for pair binding, above a threshold value of 
$U/t \approx 3$.
They also found that this attraction between doped
electrons is accompanied by a violation of Hund's rule, which requires maximal spin,
for the two-electron-doped C$_{60}$, 
and that for $U/t > 3$, the ground state for 
two-electron-doped C$_{60}$ has spin zero. ~\cite{kivelson91a, kivelson91b}
However, recent calculations,~\cite{lin05a} using quantum Monte Carlo (QMC)
techniques, suggest that the repulsive Hubbard model does not lead
to pairing on C$_{60}$. 
On the other hand, there {\it are} geometries where pair binding {\it is} known to occur~\cite{kivelson92,kivelson01}.
In particular, White \textit{et al.} in exact diagonalization (ED)
studies of the extended Hubbard model on the much smaller C$_{12}$
(truncated tetrahedron) molecule have shown that a negative
pair-binding energy (effective attraction between doped electrons)
exists for an intermediate value of the on-site Coulomb
interaction $U$ [see Eq.\ (\ref{exthubmd}) and Fig.\
\ref{edc12pbe} (a)]. A more realistic model of the fullerenes would
include longer ranged Coulomb repulsions, and it was found that
this pairing energy also survives in C$_{12}$ for
modestly repulsive values of the nearest-neighbor (NN)
interaction, $V$, but increasing $V$ eventually kills the pair
binding. 
The same violation of Hund's rule as in C$_{60}$
was also observed in C$_{12}$ [see Ref.~\onlinecite{kivelson92} and Fig.\
\ref{edc12pbe} (b)].

With a different extended Hubbard model, Sondhi \textit{et
al.}~\cite{sondhi95} studied the effects of both NN interaction $V$
and the off-diagonal interactions on the pair-binding energy and
Hund's rules violation in the C$_{60}$ molecule. Using
perturbative calculations, they find that the NN interaction $V$
terms suppress pair binding while the off-diagonal terms enhance
it. Goff and Phillips~\cite{goff92, goff93} considered the effects
of both NN interaction $V$ and longer-range terms, $V$, on the
pair-binding energy, again by perturbation theory, and also found
that the inclusion of $V$ terms strongly suppresses pair
binding in C$_{60}$.  

The fact that ED studies found pair-binding for the smaller
C$_{12}$ molecule~\cite{kivelson92} and the recent rapid development of experimental
techniques for the synthesis of C$_{20}$ solid 
phases~\cite{wang01, iqbal03} make it interesting and timely to explore correlation effects in 
C$_{20}$, the smallest gas-phase fullerene molecule
 which has dodecahedral geometry.~\cite{prinzbach00} 
In Ref.~\onlinecite{strongc20}, we briefly reported on pair-binding for
electron-doped C$_{20}$ for a wide range of values of $U/t\leq 100$, but with $V=0$, using both QMC for
$U/t\leq 3$ and ED for the full range of values.  Using cluster perturbation theory~\cite{senechal00,senechal02}
we also identified a metal-insulator transition near $U_c/t\sim 4.2$ for molecular solids formed of C$_{20}$. 
In this paper, we provide further details of our numerical techniques and consider
both electron and hole doping for an extended Hubbard model with both on-site and NN 
repulsion.
We also study density-density, spin-spin and pairing correlation functions as a 
function of separation on the molecule.

The extended Hubbard Hamiltonian on a single C$_{20}$ molecule is
defined as
\begin{equation}
H=-t\sum_{\langle
ij\rangle\sigma}(c^{\dagger}_{i\sigma}c_{j\sigma}+h.c.) +U\sum_i
n_{i\uparrow}n_{i\downarrow}+V\sum_{\langle ij\rangle}n_in_j,
\label{exthubmd}
\end{equation}
where $c^{\dagger}_{i\sigma}$ ($c_{i\sigma}$) is an electron
creation (annihilation) operator on site $i$, indices $i,j$ run
over 20 sites of a dodecahedron, $U$ is the on-site Coulomb
interaction, $V$ is the NN Coulomb interaction, and
$n_i=n_{i\uparrow}+n_{i\downarrow}$ is the number of electrons on
site $i$. 
Our goal is here to focus on strong correlation effects in C$_{20}$
using exact numerical techniques. 
The Hamiltonian
Eq.~(\ref{exthubmd}) is a simplified model of C$_{20}$ but it still largely captures such correlation effects. 
We calculate ground state energies as a function of
both $U$ and $V$ for neutral, one- and two-electron dopings.
Comparisons among these energies show that 
the electronic pair-binding energy
$\Delta_{\textrm{b}}(21)=E(20)+E(22)-2E(21)$ is positive
(repulsive) for the parameter ranges studied ($0<U/t\leq 3$ for
$V/t=0.2$ and $0.20\leq V/t\leq 0.46$ for $U/t=1$). This implies that
it is energetically favorable for two electrons to stay on 
different molecules as opposed to the same molecule. We also find that
the existence of a NN Coulomb interaction $V$ enhances this
tendency, as expected, in order to reduce the
intramolecular Coulomb interaction energy. For hole doping,
the corresponding hole pair-binding energy
$\Delta_{\textrm{b}}(19)=E(18)+E(20)-2E(19)$ is again positive
(repulsive) for the parameter range ($0<U/t\leq 3$ and $V=0$),
i.e., there is an effective repulsion between two doped holes on
the same C$_{20}$ molecule.

\begin{figure}[t]
\begin{center}
\includegraphics[clip,width=8cm]{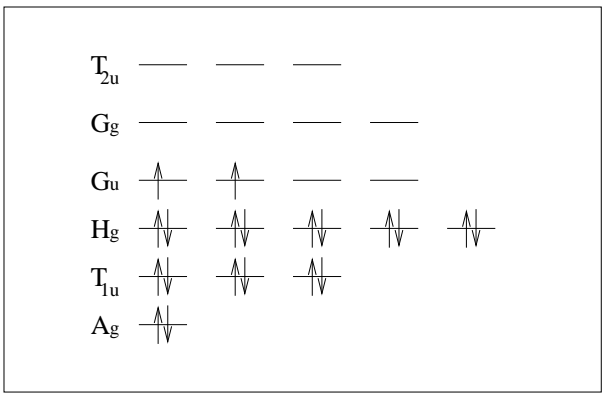}
  \caption{Huckel molecular orbitals of a neutral dodecahedral C$_{20}$ molecule. ~\cite{ellzey03}}
  \label{c20huckel}
\end{center}
\end{figure}

Unlike the case of C$_{60}$, the highest occupied molecular orbital (HOMO) of the
neutral C$_{20}$ molecules, in the weakly interacting limit, is a
four-fold orbitally degenerate level occupied by two electrons.
(See Fig.\ \ref{c20huckel}.) Hund's rules predict for this case
that the two electrons occupy different orbitals and have total
$S=1$, implying that, in the absence of a Jahn-Teller 
distortion, the neutral molecule has a magnetic
moment. In previous work~\cite{strongc20}, for $V=0$, we have confirmed this magnetic moment for $0<U/t<3$
and shown that at the metal-insulator transition, $U_c$,
the ground-state changes from a spin triplet to a singlet for
neutral C$_{20}$ and from $S=2$, through $S=1$, to $S=0$ for C$_{20}^{2-}$.
Here we extend this analysis to determine ground state spin configuration for
neutral C$_{20}$ for a fixed value of $U/t=2$ as a function of $V$, and find a level crossing between 
$V/t=1$ and $V/t=1.5$ for spin triplet and singlet states. 
For $U/t=2$, we estimate the critical $V_c/t$ to be 
$1.1$ for the spin triplet to singlet transition of the neutral 
molecule.
In light of our results for $V=0$, we expect that,
in this case too, the magnetic transition  at $V_c$ will                                                                                                   
coincide with
a metal-insulator transition for molecular solids formed
of C$_{20}$.
We also investigate the 
pair-binding energy for the hole doped case for both 
$V=0$ and $V\neq 0$, and examine the effect of a non-zero $V$ on Hund's rule.

The occurrence of orbital degeneracy and the
resulting magnetic moment are tied to the icosahedral symmetry of
the molecule. Simple molecular orbital calculations strongly
suggest that the molecular symmetry is lowered by a Jahn-Teller
effect from $I_h$ to $D_{3d}$, with the HOMO 
being a non-degenerate singlet. ~\cite{yamamoto05} However, the
correlation effects that give rise to Hund's rule compete with
this tendency to form a singlet ground state, and hence they also compete with
the Jahn-Teller effect. As reported previously,~\cite{strongc20} we find
that when the on-site Coulomb interaction $U/t$ is sufficiently large 
($U\gtrsim 4.2t$), the ground state is gapped with $S=0$ and 
the $I_h$ symmetry is likely stable against a $D_{3d}$ distortion.
In order to more exclusively focus on the effects of the non-zero $V$ term we shall here
assume that the icosahedral symmetry is unbroken even
for smaller $U$ values. 

In the next section, we briefly introduce the projection quantum
Monte Carlo (PQMC)~\cite{white89} and ED 
methods for this model. This is followed, in Section \ref{c12results}, by a 
comparison of PQMC with ED results on a C$_{12}$ and a discussion of Hund's rule
violation in C$_{12}$. In section~\ref{c20results} we focus on the C$_{20}$ 
molecule. Hole pai-rbinding in C$_{20}$ is discussed and the influence of a
non-zero next nearest neighbor $V$ on the pair-binding is investigated and results
for the triplet-singlet transition with $V$ are described along with 
calculations of several correlation functions in the C$_{20}$ 
molecule. Section \ref{conclusions} contains discussion and conclusions.

\section{Method}
\subsection{PQMC}
As noted in Ref.\ \onlinecite{zhang89}, the idea in PQMC
simulations of the extended Hubbard model is to decouple the
two-body interaction terms (both $U$ and $V$ terms) in the
partition function by means of discrete Hubbard-Stratonivich
transformations. ~\cite{hirsch83} The resultant one-body terms are
coupled to several auxiliary Ising spin fields that live either on
the lattice sites ($U$ term) or on the lattice bonds ($V$ term).
One such discrete transformation in the $V$ term is given by
\begin{equation}
e^{-\Delta\tau Vn_{i\alpha}n_{j\beta}}
=\frac{1}{2}\textrm{Tr}_{\{\sigma_{ij}^{\alpha\beta}\}}
e^{\lambda_2\sigma_{ij}^{\alpha\beta}(n_{i\alpha}-n_{j\beta})
-\frac{\Delta\tau V}{2}(n_{i\alpha}+n_{j\beta})}, \label{decouple}
\end{equation}
where $\alpha,\beta=\uparrow,\downarrow$,
$\sigma_{ij}^{\alpha\beta}=\pm 1$ is the auxiliary Ising spin on
bond $(i,j)$, $\Delta\tau$ is the discrete imaginary time slice in
PQMC, and the parameter $\lambda_2$ is determined by
$\tanh^2(\lambda_2/2)=\tanh(\frac{\Delta\tau V}{4})$. The same
decoupling equation applies for on-site Coulomb interactions,
i.e., $i=j$, except that the constant $V$ is replaced by $U$ and
$\lambda_2$ by $\lambda_1$, which is similarly given by
$\tanh^2(\lambda_1/2)=\tanh(\frac{\Delta\tau U}{4})$. These
one-body fermionic terms in the partition function can then be
explicitly traced out, leaving traces over the auxiliary
Ising spins, which can be evaluated by Monte Carlo (MC)~\cite{hirsch83}.
\begin{eqnarray}
Z&=&\sum_{\{\sigma\}}\prod_{\alpha}\det [1+B_{L}(\alpha)B_{L-1}(\alpha)\cdots
B_{1}(\alpha)]\nonumber\\
&=&\sum_{\{\sigma\}}\det O(\{\sigma\})_{\uparrow}\det O(\{\sigma\})_{\downarrow},
\label{zfinal}
\end{eqnarray}
where $\{\sigma\}=\{\sigma^1,\sigma^2,\sigma^3,\sigma^4,\sigma^5\}$ is the set
of five species of Ising fields, with $\sigma^1$ representing the on-site Ising spins and
$\sigma^{2-5}$ the NN bond Ising spins (one for each of the 4 spin configurations).
The $B_l$ matrices are defined as
\begin{eqnarray}
B_{l}(\alpha)&=&e^{-\Delta\tau K/2}e^{W^{\alpha}(l)}e^{-\Delta\tau K/2},\label{bdefine}\\
(K)_{ij}&=&\left\{\begin{array}{cc}
-t & \mbox{for $i$,$j$ NN},\\
0 & \mbox{otherwise},
\end{array}\right.\\
W_{ij}^{\alpha}(l)&=&\alpha[\delta_{ij}\lambda_1\sigma_{i}^1(l)+
\delta_{\langle ij\rangle}\lambda_2\sum_{m=2}^5\sigma_{ij}^m],\\
\delta_{\langle ij\rangle}&=&\left\{\begin{array}{cc}
1 & \mbox{for $i$,$j$ NN},\\
0 & \mbox{otherwise},
\end{array}\right.
\end{eqnarray}
where $l=1,\cdots,L$ is the time slice index, and $\alpha=\pm 1$ denotes the two
determinants in Eq.\ (\ref{zfinal}).

A complete MC sweep through the lattice will therefore consist of
trial flipping of one species of auxiliary Ising spins on all the
lattice sites and trial flipping of four species of auxiliary
Ising spins on all the NN bonds in the lattice system. Fast
calculation of the probability ratio in flipping one bond Ising
spin at one time slice is still possible using the local update
technique,~ \cite{blankenbecler81} except that one needs to apply
the probability ratio formula twice for each bond Ising spin
flip (which affects two sites).

We remark that in this decomposition scheme it is possible to
treat even longer range Coulomb interactions [e.g., next nearest
neighbor (NNN) Coulomb interactions, etc.] by introducing more
species of auxiliary Ising spins that live on these longer bonds.
The only problem is that one needs to walk through a larger and
larger phase space of the auxiliary Ising spins during the MC
simulations, which will, of course, increase the computation time.
Practically, we find that, to collect the same amount of data,
the CPU time doubles for $V\neq 0$ compared
with the $V=0$ case.

In a typical calculation the projection factor $\beta$ in PQMC was taken to be $\beta=10/t$,
and the discrete time slice was set at
$\Delta\tau=0.05/t$. $10^3$ MC warm-up sweeps through
the whole space-time lattice are typically performed before
collecting data. To estimate the statistical errors, we use the
same method as was used in Ref.\ \onlinecite{lin05a}.

\subsection{Exact Diagonalizations (ED)}
\begin{figure}[t!]
\begin{center}
\includegraphics[clip,width=8cm]{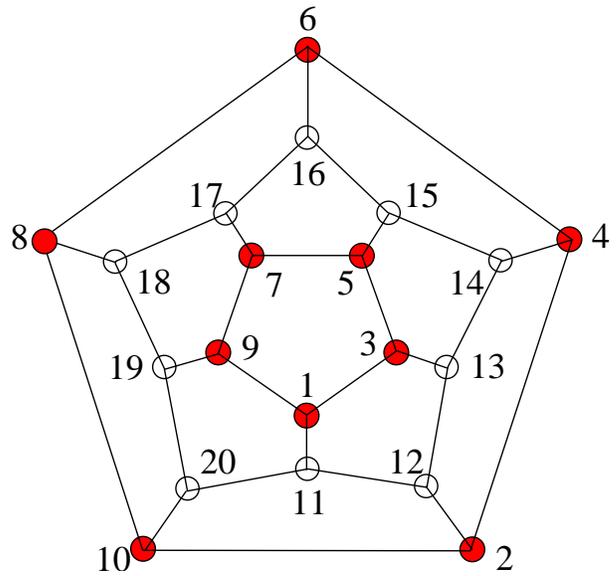}
  \caption{(Color online) Dodecahedral C$_{20}$ geometry in 2D view. Solid and empty points
  denote two sets (orbits) of carbon atoms divided by the S$_{10}$ symmetry.}
  \label{c20_real}
\end{center}
\end{figure}
The exact diagonalizations on C$_{12}$ are done using standard Lanczos techniques
and we therefore focus on the ED of C$_{20}$. We always use total particle number $N$ and
total $S_z$ as quantum numbers since they are conserved and  we perform ED in the 
corresponding reduced Hilbert space.
In addition, ED are performed using the $S_{10}$ sub-group symmetry present
in the point group $I_h$. The improper rotations generated by the elements of $S_{10}$
can be visualized as a rotation of an angle $2\pi/10$
around the center of a pentagon followed by a reflection in a plane perpendicular to the
rotation axis. This is illustrated in Fig.~\ref{c20_real} where the numbering of the sites
is to be understood in the following way: the sites 1 through 10 are shifted up by 1 (modulo 10)
under $S_{10}$ and the sites 11 through 20 are shifted in a similar manner. Hence, under 
the action of the $S_{10}$ group two different orbits exist,
marked by the solid and open points in Fig.~\ref{c20_real}. Many other symmetries
exist but the $S_{10}$ symmetry is large and relatively easy to implement, and
we have not exploited additional symmetries since the added cpu-time needed
to implement them was significant enough to offset the time gained from reducing
the size of the Hilbert space. The $S_{10}$ quantum number can be thought of as a pseudo 
angular momentum, $j_{10}$, and for each value of $N$ and $S_z$ we have to find the 
value of $j_{10}$ that corresponds to the ground-state. In many cases it is not an 
obvious value and it is often non-zero. 
In the accompanying tables we show the values of $j_{10}$ corresponding to the listed energies
and in Table~\ref{edu} we show complete dispersion of the lowest magnetic
modes for neutral C$_{20}$ as a function of $j_{10}$.

The calculations are fully parallelized Lanczos calculations executed on SHARCNET
computers. A typical calculation performed at half-filling for $N=20$, $Sz=0$ that,
after $S_{10}$ symmetry reductions, requires a Hilbert space of ${\cal N}=3,418,725,024$, is
performed with $P=64$ cpu's using about 540 seconds of cpu-time (for each cpu) per Lanczos iteration.
The memory requirement for this example is roughly 2.1Gb per cpu.
Excellent convergence is always observed with less than 300 Lanczos iterations, typically
less than 200. 

The heart of the Lanczos calculation is the matrix vector multiplication that in
this case has to be implemented in parallel. As one of several choices, we have
chosen to have each cpu apply the full matrix to one section of the vector with each
cpu returning the corresponding section of the resulting vector. The partial
results from each cpu therefore needs to be communicated between all $P$ processors with
each processor communicating to all others. Due
to the size of the involved Lanczos vectors (~40-60Gb) which greatly exceeds the available per-cpu
memory, it is necessary to repeat this $P\times P$ communication step many 
thousands of times per Lanczos step. 
The communication step therefore quickly becomes the bottle-neck in the calculation unless 
it can be done very efficiently. Fortunately, this is possible using non-blocking
communications where the individual cpu's do not wait for a communication to complete.
The draw back of using non-blocking communications is that buffer space has to be allocated 
until it has explicitly been verified that the communication has been completed. 
We have implemented a dual buffer strategy yielding an extremely efficient
communication step. The cpu-time spent per cpu is, for all accessible number
of processors we have been able to check, overwhelmingly dominated by actual calculations
rather than communications. For a fixed ${\cal N}$ we have then observed almost linear scaling
for $P=64,128,256,384$ and $512$.
The great advantage of this approach is that the complexity of Lanczos calculations
scale with the size of the Hilbert space, ${\cal N}$, as ${\cal N}\log{\cal N}$. 
Neglecting the logarithm, a doubling of the size of the Hilbert space, ${\cal N}$, can
then be almost compensated by doubling the number of processors $P$. 
\begin{table}[b]
  \centering
  \begin{tabular}{|l|cclll|}
    \hline\hline
      & $S$ & $S_z$ &  \textrm{ED} &  \textrm{PQMC} & \textrm{sign} \\
    \hline
    $E_{12}$ & 0  &   0      & -9.4647669965 & -9.466(2)  & 0.97 \\
    \hline
    $E_{13}$ & 1/2  &   1/2    & -6.8287003500 & -6.829(4)  & 0.33 \\
    \hline
    $E_{13}$ & 3/2  &   3/2    & -6.0844214907 & -6.059(6)   & 0.20 \\
    \hline
    $E_{14}$ & 0   &   0      & -4.1568425864 & -4.11(1)   & 0.11 \\
    \hline
    $E_{14}$ & 1  &   1      & -4.0772924523 & -4.080(5)  & 0.34 \\
    \hline
    $\Delta_{1,0}$   & \multicolumn{2}{l}{(1/2,0)} &  2.6360666465 & 2.637(4) &  \\
    $\Delta_{1,0}$   & \multicolumn{2}{l}{(3/2,0)} &  3.3803455058 & 3.407(6) &  \\
    $\Delta_{\textrm{b}}(13)$ &  \multicolumn{2}{l}{(0,0,1/2)} &  0.0357911171 & 0.08(1) &  \\
  \hline\hline
  \end{tabular}
  \caption{Comparison of ED and PQMC calculations
  on the truncated tetrahedron (12 sites) at $U=2t$ and $V=0.2t$.
  $E_n(S_z)$ is the energy of a system with $n$ electrons and
  $z$-component of total spin $S_z$. $\Delta_{n,m}$ is the energy difference
  $E_{12+n}(S^n_z)-E_{12+m}(S^m_z)$ with $(S^n_z,S^m_z)$ given in the
  second column. For binding energies $\Delta_{\textrm{b}}(n)$ the second column
  shows $(S^{n+1}_z,S^{n-1}_z,S^{n}_z)$ -- the $S_z$ values for 3
  states involved in its calculation~\cite{lin05a}}.\label{c12edpqmc}
\end{table}

\section{Results for C$_{12}$}\label{c12results}
Before turning our attention to the C$_{20}$ molecule we investigate the
simpler C$_{12}$ molecule in the truncated tetrahedron configuration. As mentioned above,
previous studies~\cite{kivelson92} have found a negative pair-binding energy on this
molecule that, however, became positive (repulsive interaction) in the presence of a sufficiently
large $V$. The purpose of this investigation is two-fold. First of all, we want  to verify the correctness
of our numerical approach while at the same time highlighting some of the subtleties 
of interpreting the PQMC data.   Secondly,
due to the relative ease with which calculations can be performed on this molecule it
allows for a rather detailed study of the correlation between the negative pair-binding energy
and a violation of Hund's rule for the two electron doped molecule~\cite{kivelson91a,kivelson91b}.

\subsection{Tests on the C$_{12}$ molecule}
\begin{figure}[t]
  \begin{tabular}{c}
  \includegraphics[clip,width=8cm]{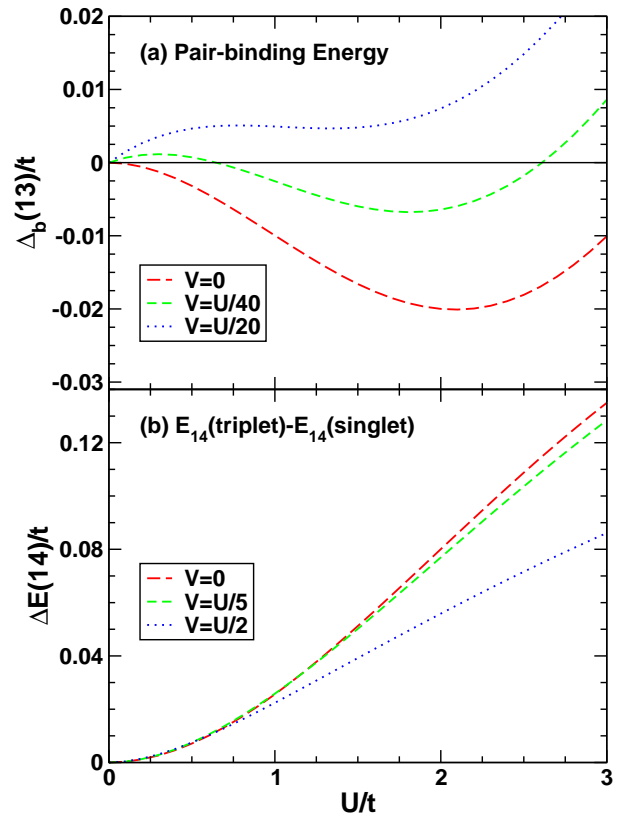} \\
  \end{tabular}
  \caption{(Color online) (a) Variation of the pair-binding energy 
  $\Delta_{\textrm{b}}(13)=E(12)+E(14)-2E(13)$ of a truncated tetrahedron molecule 
  (C$_{12}$) with $U$ and $V$ as in Fig.\ 3 in Ref.\ \onlinecite{kivelson92}. 
  (b)Hund's rules violation in the two-electron doped C$_{12}$ molecule,
  where $\Delta E(14)=E_{14}(\hbox{triplet})-E_{14}(\hbox{singlet})$.}
  \label{edc12pbe}
\end{figure}
To test our ED program, we use the same parameters as in Ref.\
\onlinecite{kivelson92} and we are able to reproduce the same pair-binding energy
as shown in Fig.\ \ref{edc12pbe} (a).
In Table\ \ref{c12edpqmc}, we see good agreement between
PQMC and ED energy values within statistical error bounds. An
exception is found for $E_{14}$ and $S_z=0$,
where the PQMC result lies a bit higher than the ED energy value.
This is due to the mixture of singlet and triplet components
in the $S_z=0$ sector and the near degeneracy of these two
states that makes the projection of the singlet ground state out
of the mixed state difficult.~\cite{lin05a} We will see that this difficulty {\it does
not} occur for C$_{20}$, where the ground state with two-electron
doping is in the spin-2 sector for $U/t\leq 3$. Hence the
pair-binding energy extracted for C$_{20}$ by PQMC for $U/t\leq 3$ is 
more reliable than the one for C$_{12}$.

\subsection{Hund's rule violation for C$_{12}^{2-}$}

In the perturbation theory studies of pair-binding in the larger fullerene 
C$_{60}$,\cite{kivelson91a,kivelson91b} it was noted that a
negative pair-binding energy (effective attraction) was correlated with
a violation of Hund's rule for the two-electron doped molecule; i.e., that
for C$_{60}^{2-}$, the ground state was found to be a singlet. 
Although our QMC results did not support the existence of pair-binding
in  C$_{60}^{2-}$ and found a spin-triplet ground state, it is of interest
to examine the correlation between pair-binding and the violation of
Hund's rule in C$_{12}$.  The non-interacting $V=U=0$ neutral molecule has completely
filled levels and hence a total spin zero. Added electrons therefore enter an unfilled
level with an orbital degeneracy of 3. Hund's rule would then predict C$_{12}^{2-}$ to
have total $S=1$. What we find is that the ground state of
C$_{12}^{2-}$ is a singlet both when the pairing is attractive and when it is
driven repulsive by increasing the nearest neighbor repulsion $V$.
This is shown in Fig.~\ref{edc12pbe}(b) where the singlet state is found
to lie below the triplet state for both positive and negative pair-binding
energies, for the range of $U$ and $V$ studied.
Thus, for this case, Hund's rule is found to be violated where pair-binding
occurs as well as where it does not.

\section{Results for C$_{20}$ molecule}\label{c20results}
\begin{figure}[t]
\begin{center}
\includegraphics[clip,width=8cm]{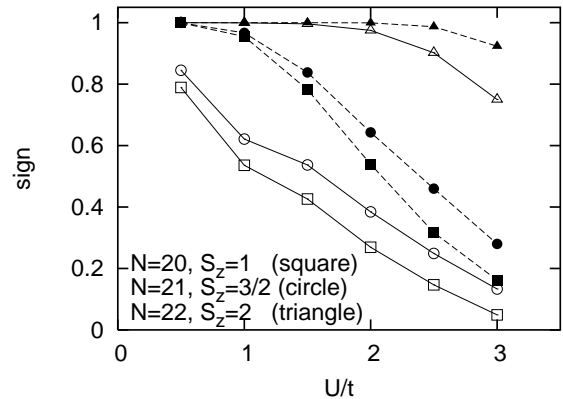}
  \caption{Average sign behavior for both $V/t=0$ (solid symbols) and
  $V/t=0.2$ (hollow symbols) at different fillings $N=20, 21, 22$.
  The lines connecting the points are guides to the eye only.}\label{signc20}
\end{center}
\end{figure}

\begin{figure}
  \begin{tabular}{c}
  \includegraphics[clip,width=8cm]{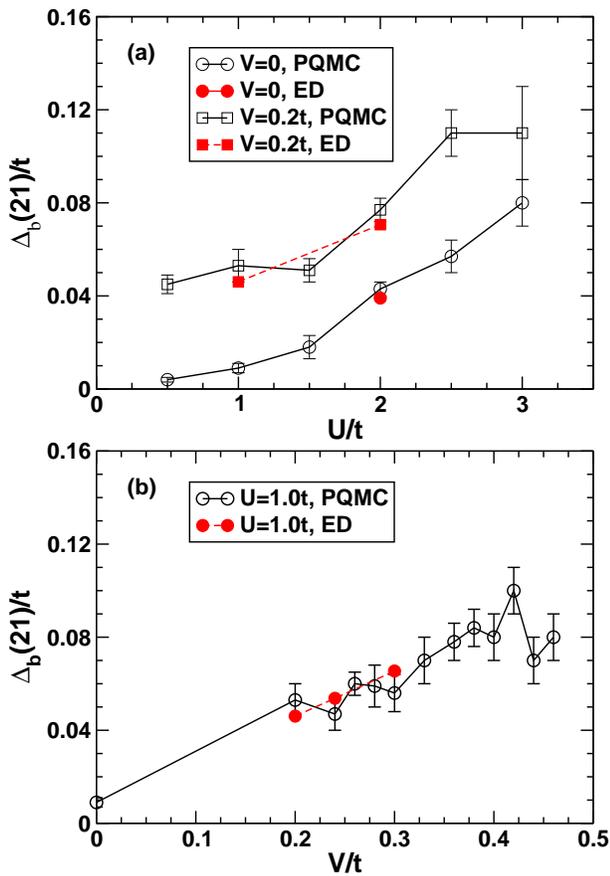} \\
  \end{tabular}
  \caption{(Color online) Electronic pair-binding energies $\Delta_{\textrm{b}}(21)/t$
  as a function of $U/t$ and $V/t$ from ED and PQMC simulations. (a) The variation of
  pair-binding energy with $U/t$ for fixed $V/t$ values. (b) The variation of
  pair-binding energy with $V/t$ for fixed $U/t=1$. The lines connecting
  MC and ED points are guides to the eye only.}
  \label{pqmc20}
\end{figure}

We now turn to the more interesting case of the C$_{20}$ molecule.
Compared with the $V=0$ case, where PQMC already has a sign problem for
the non-bipartite dodecahedral molecular geometry, the NN Coulomb
interaction $V$ terms introduce more sources of negative
probability weight, lowering the average value of the sign. Fig.\ \ref{signc20} shows the average sign for
these two cases. For the worst case ($N=20, S_z=1, U/t=3,
V/t=0.2$), where the average sign is as low as $0.05$, we have
collected $7.2\times 10^7$ MC lattice sweeps. This gives a
relatively large but nevertheless meaningful error bar.
[See Fig.\ \ref{pqmc20} (a).] For
other parameter values, we have collected about $2.2\times 10^7$
MC sweeps. The acceptance ratio for the on-site Ising spin trial
flipping ranges from $80\%$ ($U/t=3$) to $93\%$ ($U/t=1$), while
that for the bond Ising spins is about $95\%$ due to the small
value of $V/t=0.2$.

\begin{table}[b]
  \centering
  \begin{tabular}{|l|c|c|lc|ll|}
    \hline\hline
      & $S$ & $S_z$ &  \textrm{ED} & $j_{10}$ & \textrm{PQMC} & \textrm{sign} \\
    \hline
    $E_{18}$ & 0  &   0      & -22.4044466933 & 0  & -22.402(1)  & 1.00 \\
    $E_{18}$ & 1  &   1      & -21.6778357505 & $\pm 3,5$  &-21.637(2)  & 0.49 \\
    \hline
    $E_{19}$ & 1/2  &   1/2  & -21.5223243600 & $\pm 1,\pm$ 3 &-21.5227(6)  & 0.64 \\
    $E_{19}$ & 3/2  &   3/2  & -20.8990191757 & 0,$\pm 4$ &-20.826(3)  & 0.35 \\
    \hline
    $E_{20}$ & 1  &   0      & -20.5983834340 & $0,\pm 2$& -20.533(3)  & 0.26 \\
    $E_{20}$ & 1   &   1      & -20.5983834340 &$0,\pm 2$&  -20.597(2)  & 0.54 \\
    $E_{20}$ & 0   &   0      & -20.5920234654 &$0,\pm 2, \pm 4 $ &  & \\
    $E_{20}$ & 2   &   2      & -19.9634427212 &$\pm 2, \pm 4, 5$ &  & \\
    \hline
    $E_{21}$ & 3/2  &   1/2    & -19.6331786587 &$ \pm 1, \pm 3$ &-19.465(8)  & 0.19 \\
    $E_{21}$ & 3/2  &   3/2    & -19.6331786587 &$ \pm 1, \pm 3$ &-19.634(1)  & 0.64 \\
    \hline
    $E_{22}$ & 2  &   0      & -18.6289129089 & 0  & -18.282(7)  & 0.10 \\
    $E_{22}$ & 2  &   1      & -18.6289129089 & 0  & -18.448(5)  & 0.32 \\
    $E_{22}$ & 2  &   2      & -18.6289129089 & 0  & -18.628(1)  & 1.00 \\
  \hline\hline
  \end{tabular}
\caption{Comparison of ground state energies from ED and PQMC calculations
on the C$_{20}$ molecule at $U/t=2$ and $V=0$. 
See the caption in Table\ \ref{c12edpqmc}
for the corresponding definition of various quantities.} \label{c20edpqmc}
\end{table}

\begin{figure}[t]
\begin{center}
\includegraphics[clip,width=8cm]{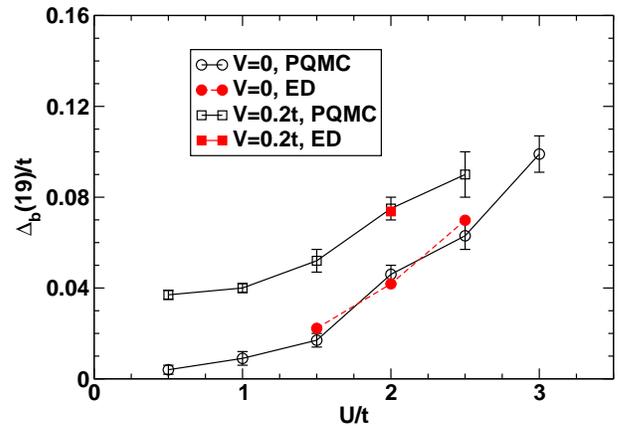}
  \caption{(Color online) Hole pair-binding energies $\Delta_{\textrm{b}}(19)/t$
  as a function of $U/t$ for $V=0$ from ED and PQMC simulations. 
  The lines connecting
  MC points are guides to the eye only.}
  \label{pqmc20h}
\end{center}
\end{figure}
\begin{table}[b]
  \centering
  \begin{tabular}{|l|ccl|cclc|}
    \hline\hline
           & $S$ & $S_z$    &  $U/t=3$  &$S$ & $S_z$    &  $U/t=5$(ED)   & $j_{10}$ \\
    \hline
    $E_{20}$  & 1   &   0      &   -17.04(2)  & 0 & 0 &  -12.111284292 & 5\\
    $E_{20}$  & 1   &   1      &   -17.036(6) & 1 & 1 &  -11.877033283 & $0,\pm 2$\\
    $E_{21}$  & 3/2 &   1/2    &   -15.29(6)  & 1/2 & 1/2 &  -9.1165560273 & $\pm 1, \pm 3, 5$ \\
    $E_{21}$  & 3/2 &   3/2    &   -15.529(5) & 3/2 & 3/2 &  -8.9633623599 & $\pm 1, \pm 3$    \\
    $E_{22}$  & 2   &   0      &   -13.936353 & 1 & 0 &  -5.9715313615  & $\pm 2, \pm 4$\\
    $E_{22}$  & 2   &   1      &   -13.81(2)  & 1 & 1 &   -5.9715313615 & $\pm 2, \pm 4$\\
    $E_{22}$  & 2   &   2      &   -13.935(1) &   &   &    & \\
  \hline\hline
  \end{tabular}
  \caption{Ground state energies for neutral, one- and two-electron-doped C$_{20}$
  molecules at $U/t=3, 5$ and $V=0$ from PQMC and ED, which shows a transition between Hund's and
  anti-Hund's states at $3<U/t<5$ for neutral, one-, and two-electron-doped molecules, respectively.
  Data without error bars are from ED.} \label{antihund}
\end{table}

\begin{table*}
  \centering
  \begin{tabular}{|cc|cccccc|}
    \hline\hline
      $U$ & $S$ & $j_{10}=0$ & $j_{10}=\pm 1$ & $j_{10}=\pm 2$ & $j_{10}=\pm 3$ & $j_{10}=\pm 4$ & $j_{10}=5$ \\
    \hline
    &                         $ 0$ & -20.5920234655  &     $>$ -19.90               & -20.5920234655 &   $>$ -19.90              & -20.5920234655 &  -20.0527029539\\
   \raisebox{1.5ex}[0pt]{2} & $ 1$ & -20.5983834340  &   -19.9776970001   & -20.5983834340 & -19.9776970001  &
-20.5981592741 &  -19.9634427213\\
    \hline
          &                    $ 0$ & -12.0123014488  &   -11.6726562451   & -12.0123014488 & -11.6726562451  & -12.0123014488 &  -12.1112842959\\
   \raisebox{1.5ex}[0pt]{5} & $ 1$ & -11.8770332831  &   -11.8472120431   & -11.8770332831 & -11.8472120431  &
-11.8103044760 &  -11.8118179567\\
    \hline
          &                    $ 0$ &-8.0452584717  &   -7.806831   &     -8.0452584717 &  -7.8068365859  & -8.0452584717 &  -8.1803385740\\
   \raisebox{1.5ex}[0pt]{8} & $ 1$ & -7.9497836200  &   -7.9415479844   & -7.9497836200 &  -7.9415479844  & -7.8490047592 &  -7.9156714009\\
  \hline\hline
  \end{tabular}
  \caption{ED results for the dispersion of the lowest singlet and triplet states with $j_{10}$ for neutral C$_{20}$ with $U/t=2, 5$ and $8$ in
    all cases with $V/t=0$.
  } \label{edu}
\end{table*}

\subsection{Pair-binding energy}
Table\ \ref{c20edpqmc} shows the energies of the C$_{20}$ molecule
at different fillings from PQMC and ED for $U/t=2$ and $V=0$. Both ED and PQMC 
predict the ground states to be in the same spin sectors for the molecule, 
and the calculated energies are in agreement within MC
error bounds.

In order to understand the comparison of PQMC and ED data in Table\ \ref{c20edpqmc},
it is important to recognize a systematic weakness of PQMC which is that, when the
ground state is a spin multiplet, the different partners appear to have different
energies, increasing with decreasing values of $|S_z|$, because the states with
smaller values of $|S_z|$ mix with higher lying states that have the same value
of $|S_z|$.  In general except for statistical error, a state with $S_z=0$ will
appear to lie above its partners with the same total $S$.  This tendency is apparent
in the results for $E_{20}$ ($S$=1), $E_{21}$ ($S$=3/2), and $E_{22}$ ($S$=2).
Conversely, if a ground state with $S_z=0$ lies below a state with $S_z=1$, we
expect the ground state to be a singlet.  However, in this case, the value of
the ground state energy will be perturbed upward by any admixture of the next higher
state with $S=1, S_z=0$, as happened for $E_{14}(S=0)$ in Table\ \ref{c12edpqmc}.
In general it is also true that accurate PQMC results are more easily obtained when
the average sign is close to 1 compared to when the average sign is small.

Pair-binding energies $\Delta_{\textrm{b}}(21)/t$ (electron)
and $\Delta_{\textrm{b}}(19)/t$ (hole)  as a
function of both $U/t$ and $V/t$ are shown in  Fig.\ \ref{pqmc20} and
\ref{pqmc20h}, respectively. For $V=0$, for both electron and hole
doping, we see that the pair-binding energy is always positive
(repulsive) for $U/t>0$, and increases with increasing $U/t$. This
is the same behavior as we observed for the C$_{60}$
molecule.~\cite{lin05a} Turning on the NN Coulomb interaction $V$
($V/t=0.2$ in Fig.\ \ref{pqmc20}(a)) increases the
pair-binding energy further. Hence, putting two extra
electrons on the same neutral molecule becomes more costly when
the NN Coulomb interaction is not negligible. 
 Panel (b) in Fig.\ \ref{pqmc20}
shows the variation of the pair-binding energy as a function of
$V/t$ for fixed $U/t=1$. Again the pair-binding energy is positive
(repulsive), and generally increases with $V/t$. The agreement
between ED and PQMC results is fairly good and even though the PQMC
data show some tendency to non-monotonic behavior for this interval
of $V/t$, the ED results show that this is explained by the natural
statistical spread of the data.
Hence, in the regime $V<V_c,\ U<U_c$, the pair-binding energy increases 
with both $U$ and $V$ and energetically, it becomes increasingly favorable for two electrons to
stay on two different C$_{20}$ molecules. However, we note that, for $V=0,\ U>U_c$ it was previously found~\cite{strongc20}
that the pair-binding energy {\it decreases} with $U$, reaching a minimum at $U/t\sim 10$, before increasing
and reaching a finite value in the $U\to\infty$ limit.

\subsection{Hund's rule}
It is also clear, from the data in Table \ref{c20edpqmc} and \ref{antihund}, that Hund's rule is 
obeyed for the corresponding range of parameters, i.e., $U/t\leq3$, $V=0$. That is, the ground 
state for 20 through 22 all have the maximum values of total spin for electrons outside
the $C_{20}^{2+}$ core, ranging from total spin 1 for 
20 electrons through total spin 2 for 22 electrons. This behavior occurs in the range of 
parameters where PQMC converges (for maximal $|S_z|$ as discussed above.)
As $U/t$ is increased above 3, the sign problem prevents 
reliable PQMC calculations. This difficulty does not arise 
in ED where accurate calculations are possible for essentially any value of $U/t$. We have used 
ED to explore what happens for larger values of $U/t$. \cite{strongc20} For example, results 
for $U=5t$ are shown in the right hand columns of Table\ \ref{antihund}. Here Hund's rule is 
clearly violated. For 20 electrons, the ground state has spin zero; for 21 electrons the ground 
state has spin 1/2; while for 22 electrons the ground state has spin 1. Clearly there are level 
crossings in the range $3 < U/t < 5$. Additional results in this regime are given in 
Ref.\ \onlinecite{strongc20}. ED also allows the calculation of the spin gap, the gap between 
the ground state and the lowest lying excited state with different total spin. Results are 
shown in Table\ \ref{edu} for a neutral C$_{20}$ molecule with $U = 2,5,8$ and $V=0$. When the 
metal-insulator transition occurs in the vicinity of $U_c/t\sim 4.2$, the ground-state spin changes 
from an orbitally degenerate $S=1$ for $U<U_c$ to a non-degenerate singlet for $U>U_c$. From the 
results presented in table~\ref{edu} we see that it is the singlet state at $j_{10}=5$ that moves 
toward the bottom of the spectrum with increasing $U$ and eventually, for $U>U_c$ becomes the ground-state.
Focusing on the case $V/t=0$, we see from table~\ref{edu} that at $U/t=2$, the ground-state energy 
is a singlet $E^{1}/t=-20.5983834340$ with a gap to the lowest
lying singlet of $\Delta E^{1,0}/t = 0.0063599685$.
Here the superscripts denote the spin of the ground- and excited states, respectively.
For $U/t\geq 5$ we find that the ground-state for the $I_h$ configuration now
is a non-degenerate singlet, $S=0$, with energy
$E^{0}/t=-12.1112842922$. The lowest lying 
triplet excitation with $\Delta E^{0,1}/t =0.2342510092$. This picture continues to hold for larger $U/t$ with the
triplet gap at $U/t=8$ only slightly larger, $\Delta E^{0,1}/t=0.2305549540$.

Next we explore the ground state spin of the neutral molecule with different $V/t$ values for a fixed 
$U/t=2$. Using ED techniques we determine that the ground-state for $V/t=1$ and $V/t=1.5$ in both
cases occur for $j_{10}=0$. However,
the ground state changes from a spin triplet for $V/t=1$ to a spin singlet for $V/t=1.5$. Specifically, we find at
$V/t=1$, E(Singlet)=$5.702018$ and E(triplet)=$5.639496$, whereas for $V/t=1.5$ we find
E(Singlet)=$17.318536$ and E(triplet)=$17.499741$.
By assuming a linear 
dependence of the energy on $V/t$ in this region, we determine that the level crossing 
occurs near $V_c/t\sim 1.1$ for $U/t=2$.

\subsection{Correlation functions}
We have also investigated what other correlations might be induced in the C$_{20}$ molecule by
calculating the following correlation functions: charge-charge, spin-spin, and pairing 
correlations as a function of lattice distance. Similar calculations for the C$_{60}$ molecule have been 
reported in Refs.\ \onlinecite{scalettar93} and \onlinecite{rokhsar93}. 
\begin{figure}[t]
\begin{center}
\includegraphics[clip,width=8cm]{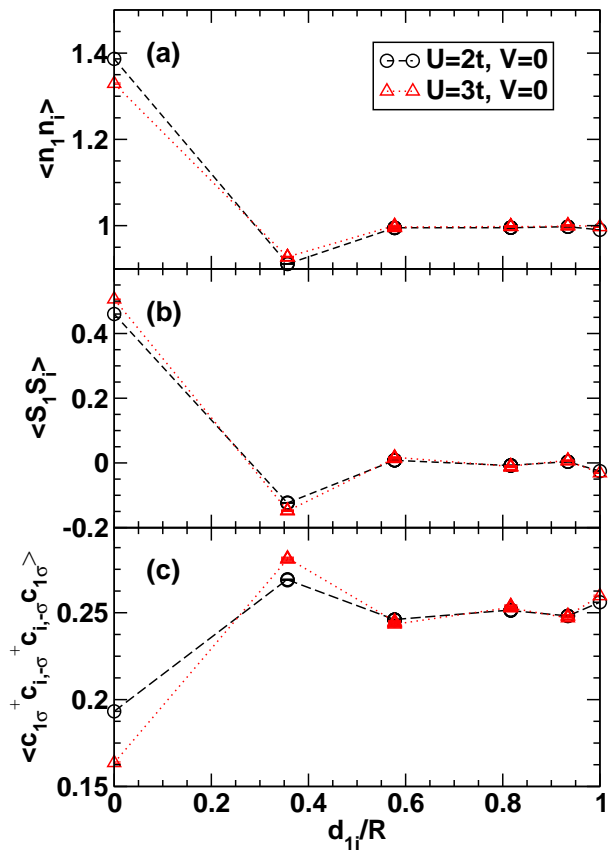}
  \caption{(Color online) Variation of (a) charge-charge, (b) spin-spin, and (c) pairing
correlation functions for $U=2t, 3t$ and $V=0$ for a C$_{20}$
molecule with respect to the lattice site spacing. $d_{1i}$ is distance
between site 1 and $i$. $R$ is diameter of C$_{20}$ molecule.}
  \label{corr}
\end{center}
\end{figure}
                                                                                                               
\begin{figure}[t]
\begin{center}
\includegraphics[clip,width=8cm]{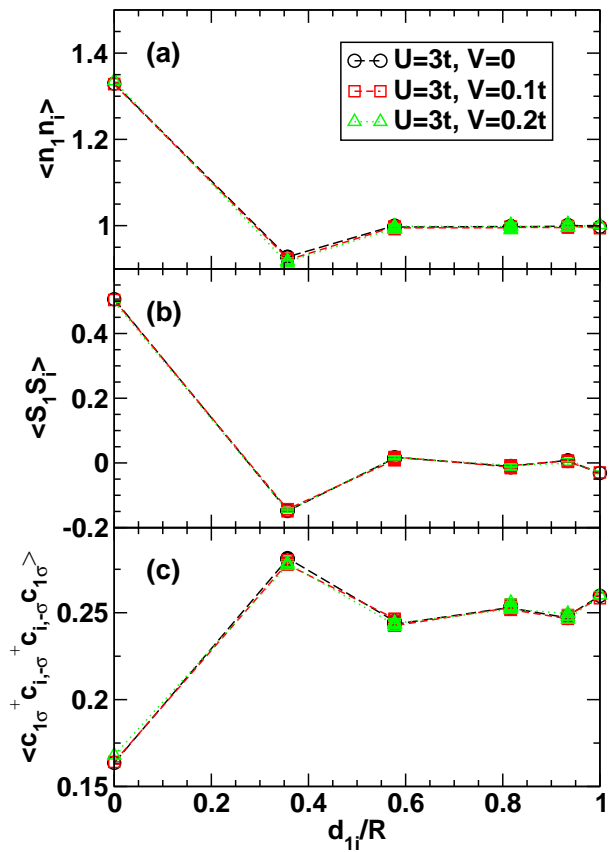}
  \caption{(Color online) Variation of (a) charge-charge, (b) spin-spin, and (c) pairing
correlation functions for $U=3t$ and $V=0, 0.1t, 0.2t$ for a
C$_{20}$ molecule with respect to the lattice site spacing.}
  \label{corrV}
\end{center}
\end{figure}

We define the correlation functions with respect to 
lattice site 1 in the neutral molecule: $\langle n_1n_i\rangle$ is the charge-charge correlation,
$\langle S_1\cdot S_i\rangle$ is the spin-spin correlation, and 
$\langle c_{1\sigma}^{\dagger}c_{i,-\sigma}^{\dagger}c_{i,-\sigma}c_{1,\sigma}\rangle$ is the 
pairing correlation, where $i=1,\ldots,20$. In 
Fig.\ \ref{corr}, we show the variation of these correlations for $U=2t$ and $3t$ as a function 
of lattice spacing $d_{1i}/R$, where $d_{1i}$ is the distance between site 1 and $i$, and $R$ 
is the molecular diameter. For the dodecahedral geometry, there are only 5 inequivalent
neighbors, all at distinct distances.  
One can understand the on-site correlations in terms of the 
probabilities, $p_n$, $n$=0,1,2, for having $n$ electrons on each site.  Then, the on-site
correlations functions are $\langle n_1^2\rangle=p_1+4p_2$, $\langle S_1^2\rangle=3p_1/4$,
and $\langle c_{1\sigma}^{\dagger}c_{1,-\sigma}^{\dagger}c_{1,-\sigma}c_{1,\sigma}\rangle=p_2.$

In Fig.~\ref{corr}(a) we show results for the charge-charge correlation function for 2 different
values of $U$ with $V=0$.
As expected, the on-site charge-charge correlation is reduced by an
increase of the on-site Coulomb interaction $U$ [panel (a)]. At larger distances, the charge 
on site 1 and $i$ are uncorrelated. The unit value of the charge-charge correlation corresponds to 
uniform distribution of charge. 

Fig.~\ref{corr}(b) shows the spin-spin correlation function again for $U/t=2,3$ with $V=0$. The NN spin-spin 
correlation has a negative finite value, and its magnitude is enhanced by a larger $U$ value. 
For spatial distances larger than 1 we see that this correlation function quickly approaches 0.
Similar behavior has been observed for the spin-spin correlation in the C$_{60}$ molecule 
in Ref.~\onlinecite{scalettar93}
and it was suggested that the rapid decay of the spin-spin
correlation function was indicative of a resonant valence bond (RVB) or ``spin dimer'' 
state. The similarity between our results and those of Ref.~\onlinecite{scalettar93}, suggest that
the spin correlations in the ground-state of C$_{20}$ also might be described by considering valence bond states including only
dimers of relatively short length. 

QMC results for the pair correlation 
are shown in Fig.~\ref{corr}(c) with $U/t=2,3$ and $V=0$.
Interestingly, there
is a peak of the pairing order when site 1 and $i$ are NN sites. This again supports
the RVB or ``spin dimer'' model for the ground-state. 
Beyond the nearest-neighbor distance, the pairing correlation function, along with the other correlation functions, is very close to its uncorrelated value 
except for $d_{1i}=R$, where the pairing order parameter is slightly enhanced. At the same time
the spin-spin correlation is slightly negative showing
an antiferromagnet correlation. 
This corresponds to the ``dumb bell'' model proposed in Ref.~\onlinecite{rokhsar93} where electron pairs are
formed at the maximal distances of the molecular diameter. We note that in the present case, the enhancement of
the correlations at the distances of $R$ corresponding to 
this ``dumb bell'' 
pairing is relatively weak. 

We have also studied the influence of a non-zero $V$ on the correlations. In Fig.~\ref{corrV} we show
results for a fixed $U/t=3$ and three different values of $V=0,0.1t, 0.2t$. Clearly, the effect of the NN Coulomb interaction 
$V$ on these correlation functions is relatively weak, with the curves being almost identical for the range of $V$ 
considered here.

\section{Conclusions}\label{conclusions}
In this paper we have studied the extended Hubbard model on a
C$_{20}$ molecule through ED and PQMC simulations. The comparison clearly
elucidates the relative strengths of the two methods.  PQMC is possible for much larger
systems than can be treated by ED. However, ED has been applied successfully to 
the Hubbard model on 20 sites with
18-22 electrons, by making 
effective use of the capabilities of a large number of coupled processors. 
PQMC works best when the ground state is well separated from excited states
with the same value of $S_z$.  As a result, ground states with larger total
spin $S$ and maximal $|S_z|$ are most accurately determined, while ground
states with $S=0$ are sometimes problematic.  This behavior was also found
in our earlier work on $C_{60}$,\cite{lin05a} and the comparison of ED
and PQMC results for $C_{20}$ is consistent with and lends confidence
to those earlier results.

The
pair-binding energy  for $C_{20}$ shows that extra added electrons (holes)
prefer to sit on different molecules, rather than to reside in
pairs on molecules. This rules out the possibility that the
extended Hubbard model on a single C$_{20}$ molecule can produce
an effective attraction between electrons (holes) from purely
electronic interactions. Our earlier work showed that this conclusion applies to
the C$_{60}$ molecule as well.~\cite{lin05a} We also
find that Hund's rule is obeyed for $U/t \le 3$ and small values of $V$
and that larger values of $U$ and $V$ lead to level crossings and ground states
for which Hund's rule is violated. For fixed $V=0$, we
have determined that this transition happens between $U/t=3$ and
$U/t=5$, at $U_c/t\sim4.2$. And for fixed $U/t=2$, as a function of $V$, we have determined that this transition happens between 
$V/t=1$ and $V/t=1.5$, at $V_c\sim 1.1$. As was the case at the transition occurring at $U_c/t\sim 4.2$ for $V=0$, we expect
this transition to coincide with a metal-insulator transition for molecular solids formed of C$_{20}$. 
More generally, for $U/t\leq3$ and $V/t\leq 0.2$, we find that the spin, charge and pairing
correlations fall off rapidly even in the presence of NN Coulomb repulsion. It is an interesting
open question if molecular solids formed of C$_{20}$, in particular away from half-filling, would
display non-trivial order for $V>V_c$. The answer to this question would be numerically demanding
and we have therefore left it for future work.

\begin{acknowledgments}
This project was supported by the Natural Sciences and Engineering
Research Council of Canada, the Canadian Institute for
Advanced Research, and the Canadian Foundation for Innovation.
FL is supported by the US Department of Energy under award number
DE-FG52-06NA26170.  AJB, CK and ESS gratefully acknowledge the hospitality
of the Kavli Institute for Theoretical Physics in Santa
Barbara, where part of this work was carried out and supported by
the NSF under Grant No. PHY05-51164.
 All the calculations were performed using  SHARCNET
supercomputing facilities.
\end{acknowledgments}

\bibliography{exthub}

\begin{thebibliography}{22}
\expandafter\ifx\csname natexlab\endcsname\relax\def\natexlab#1{#1}\fi
\expandafter\ifx\csname bibnamefont\endcsname\relax
  \def\bibnamefont#1{#1}\fi
\expandafter\ifx\csname bibfnamefont\endcsname\relax
  \def\bibfnamefont#1{#1}\fi
\expandafter\ifx\csname citenamefont\endcsname\relax
  \def\citenamefont#1{#1}\fi
\expandafter\ifx\csname url\endcsname\relax
  \def\url#1{\texttt{#1}}\fi
\expandafter\ifx\csname urlprefix\endcsname\relax\def\urlprefix{URL }\fi
\providecommand{\bibinfo}[2]{#2}
\providecommand{\eprint}[2][]{\url{#2}}

\bibitem[{\citenamefont{Chakravarty et~al.}(1991)\citenamefont{Chakravarty,
  Gelfand, and Kivelson}}]{kivelson91a}
\bibinfo{author}{\bibfnamefont{S.}~\bibnamefont{Chakravarty}},
  \bibinfo{author}{\bibfnamefont{M.~P.} \bibnamefont{Gelfand}},
  \bibnamefont{and} \bibinfo{author}{\bibfnamefont{S.}~\bibnamefont{Kivelson}},
  \bibinfo{journal}{Science} \textbf{\bibinfo{volume}{254}},
  \bibinfo{pages}{970} (\bibinfo{year}{1991}).

\bibitem[{\citenamefont{Chakravarty and Kivelson}(1991)}]{kivelson91b}
\bibinfo{author}{\bibfnamefont{S.}~\bibnamefont{Chakravarty}} \bibnamefont{and}
  \bibinfo{author}{\bibfnamefont{S.}~\bibnamefont{Kivelson}},
  \bibinfo{journal}{Europhys. Lett.} \textbf{\bibinfo{volume}{16}},
  \bibinfo{pages}{751} (\bibinfo{year}{1991}).

\bibitem[{\citenamefont{Chakravarty and Kivelson}(2001)}]{kivelson01}
\bibinfo{author}{\bibfnamefont{S.}~\bibnamefont{Chakravarty}} \bibnamefont{and}
  \bibinfo{author}{\bibfnamefont{S.}~\bibnamefont{Kivelson}},
  \bibinfo{journal}{\prb} \textbf{\bibinfo{volume}{64}},
  \bibinfo{pages}{064511} (\bibinfo{year}{2001}).

\bibitem[{\citenamefont{Lin et~al.}(2005)\citenamefont{Lin, \v{S}makov,
  S\o{}rensen, Kallin, and Berlinsky}}]{lin05a}
\bibinfo{author}{\bibfnamefont{F.}~\bibnamefont{Lin}},
  \bibinfo{author}{\bibfnamefont{J.}~\bibnamefont{\v{S}makov}},
  \bibinfo{author}{\bibfnamefont{E.~S.} \bibnamefont{S\o{}rensen}},
  \bibinfo{author}{\bibfnamefont{C.}~\bibnamefont{Kallin}}, \bibnamefont{and}
  \bibinfo{author}{\bibfnamefont{A.~J.} \bibnamefont{Berlinsky}},
  \bibinfo{journal}{Phys. Rev. B} \textbf{\bibinfo{volume}{71}},
  \bibinfo{pages}{165436} (\bibinfo{year}{2005}).

\bibitem[{\citenamefont{White et~al.}(1992)\citenamefont{White, Chakravarty,
  Gelfand, and Kivelson}}]{kivelson92}
\bibinfo{author}{\bibfnamefont{S.~R.} \bibnamefont{White}},
  \bibinfo{author}{\bibfnamefont{S.}~\bibnamefont{Chakravarty}},
  \bibinfo{author}{\bibfnamefont{M.~P.} \bibnamefont{Gelfand}},
  \bibnamefont{and} \bibinfo{author}{\bibfnamefont{S.~A.}
  \bibnamefont{Kivelson}}, \bibinfo{journal}{Phys. Rev. B}
  \textbf{\bibinfo{volume}{45}}, \bibinfo{pages}{5062} (\bibinfo{year}{1992}).

\bibitem[{\citenamefont{Sondhi et~al.}(1995)\citenamefont{Sondhi, Gelfand, Lin,
  and Campbell}}]{sondhi95}
\bibinfo{author}{\bibfnamefont{S.~L.} \bibnamefont{Sondhi}},
  \bibinfo{author}{\bibfnamefont{M.~P.} \bibnamefont{Gelfand}},
  \bibinfo{author}{\bibfnamefont{H.~Q.} \bibnamefont{Lin}}, \bibnamefont{and}
  \bibinfo{author}{\bibfnamefont{D.~K.} \bibnamefont{Campbell}},
  \bibinfo{journal}{Phys. Rev. B} \textbf{\bibinfo{volume}{51}},
  \bibinfo{pages}{5943} (\bibinfo{year}{1995}).

\bibitem[{\citenamefont{Goff and Phillips}(1992)}]{goff92}
\bibinfo{author}{\bibfnamefont{W.~E.} \bibnamefont{Goff}} \bibnamefont{and}
  \bibinfo{author}{\bibfnamefont{P.}~\bibnamefont{Phillips}},
  \bibinfo{journal}{Phys. Rev. B} \textbf{\bibinfo{volume}{R46}},
  \bibinfo{pages}{603} (\bibinfo{year}{1992}).

\bibitem[{\citenamefont{Goff and Phillips}(1993)}]{goff93}
\bibinfo{author}{\bibfnamefont{W.~E.} \bibnamefont{Goff}} \bibnamefont{and}
  \bibinfo{author}{\bibfnamefont{P.}~\bibnamefont{Phillips}},
  \bibinfo{journal}{Phys. Rev. B} \textbf{\bibinfo{volume}{48}},
  \bibinfo{pages}{3491} (\bibinfo{year}{1993}).

\bibitem[{\citenamefont{Wang et~al.}(2001)\citenamefont{Wang, Ke, Zhu, Zhu,
  Ruan, Chen, Huang, and Zheng}}]{wang01}
\bibinfo{author}{\bibfnamefont{Z.}~\bibnamefont{Wang}},
  \bibinfo{author}{\bibfnamefont{X.}~\bibnamefont{Ke}},
  \bibinfo{author}{\bibfnamefont{Z.}~\bibnamefont{Zhu}},
  \bibinfo{author}{\bibfnamefont{F.}~\bibnamefont{Zhu}},
  \bibinfo{author}{\bibfnamefont{M.}~\bibnamefont{Ruan}},
  \bibinfo{author}{\bibfnamefont{H.}~\bibnamefont{Chen}},
  \bibinfo{author}{\bibfnamefont{R.}~\bibnamefont{Huang}}, \bibnamefont{and}
  \bibinfo{author}{\bibfnamefont{L.}~\bibnamefont{Zheng}},
  \bibinfo{journal}{Phys. Lett. A} \textbf{\bibinfo{volume}{280}},
  \bibinfo{pages}{351} (\bibinfo{year}{2001}).

\bibitem[{\citenamefont{Iqbal et~al.}(2003)\citenamefont{Iqbal, Zhang, Grebel,
  Vijayalakshmi, Lahamer, Benedek, Bernasconi, Cariboni, Spagnolatti, Sharma
  et~al.}}]{iqbal03}
\bibinfo{author}{\bibfnamefont{Z.}~\bibnamefont{Iqbal}},
  \bibinfo{author}{\bibfnamefont{Y.}~\bibnamefont{Zhang}},
  \bibinfo{author}{\bibfnamefont{H.}~\bibnamefont{Grebel}},
  \bibinfo{author}{\bibfnamefont{S.}~\bibnamefont{Vijayalakshmi}},
  \bibinfo{author}{\bibfnamefont{A.}~\bibnamefont{Lahamer}},
  \bibinfo{author}{\bibfnamefont{G.}~\bibnamefont{Benedek}},
  \bibinfo{author}{\bibfnamefont{M.}~\bibnamefont{Bernasconi}},
  \bibinfo{author}{\bibfnamefont{J.}~\bibnamefont{Cariboni}},
  \bibinfo{author}{\bibfnamefont{I.}~\bibnamefont{Spagnolatti}},
  \bibinfo{author}{\bibfnamefont{R.}~\bibnamefont{Sharma}},
  \bibnamefont{et~al.}, \bibinfo{journal}{Eur. Phys. J. B}
  \textbf{\bibinfo{volume}{31}}, \bibinfo{pages}{509} (\bibinfo{year}{2003}).

\bibitem[{\citenamefont{Prinzbach et~al.}(2000)\citenamefont{Prinzbach, Weller,
  Landenberger, Wahl, Worth, Scott, Gelmont, Olevano, and van
  Issendorff}}]{prinzbach00}
\bibinfo{author}{\bibfnamefont{H.}~\bibnamefont{Prinzbach}},
  \bibinfo{author}{\bibfnamefont{A.}~\bibnamefont{Weller}},
  \bibinfo{author}{\bibfnamefont{P.}~\bibnamefont{Landenberger}},
  \bibinfo{author}{\bibfnamefont{F.}~\bibnamefont{Wahl}},
  \bibinfo{author}{\bibfnamefont{J.}~\bibnamefont{Worth}},
  \bibinfo{author}{\bibfnamefont{L.~T.} \bibnamefont{Scott}},
  \bibinfo{author}{\bibfnamefont{M.}~\bibnamefont{Gelmont}},
  \bibinfo{author}{\bibfnamefont{D.}~\bibnamefont{Olevano}}, \bibnamefont{and}
  \bibinfo{author}{\bibfnamefont{B.}~\bibnamefont{van Issendorff}},
  \bibinfo{journal}{Nature} \textbf{\bibinfo{volume}{407}}, \bibinfo{pages}{60}
  (\bibinfo{year}{2000}).

\bibitem[{\citenamefont{Lin et~al.}(2007)\citenamefont{Lin, S\o{}rensen,
  Kallin, and Berlinsky}}]{strongc20}
\bibinfo{author}{\bibfnamefont{F.}~\bibnamefont{Lin}},
  \bibinfo{author}{\bibfnamefont{E.~S.} \bibnamefont{S\o{}rensen}},
  \bibinfo{author}{\bibfnamefont{C.}~\bibnamefont{Kallin}}, \bibnamefont{and}
  \bibinfo{author}{\bibfnamefont{A.~J.} \bibnamefont{Berlinsky}}
  (\bibinfo{year}{2007}), \bibinfo{note}{cond-mat/0701727}.

\bibitem[{\citenamefont{S\'en\'echal et~al.}(2000)\citenamefont{S\'en\'echal,
  Perez, and Pioro-Ladri\`ere}}]{senechal00}
\bibinfo{author}{\bibfnamefont{D.}~\bibnamefont{S\'en\'echal}},
  \bibinfo{author}{\bibfnamefont{D.}~\bibnamefont{Perez}}, \bibnamefont{and}
  \bibinfo{author}{\bibfnamefont{M.}~\bibnamefont{Pioro-Ladri\`ere}},
  \bibinfo{journal}{\prl} \textbf{\bibinfo{volume}{84}}, \bibinfo{pages}{522}
  (\bibinfo{year}{2000}).

\bibitem[{\citenamefont{S\'en\'echal et~al.}(2002)\citenamefont{S\'en\'echal,
  Perez, and Plouffe}}]{senechal02}
\bibinfo{author}{\bibfnamefont{D.}~\bibnamefont{S\'en\'echal}},
  \bibinfo{author}{\bibfnamefont{D.}~\bibnamefont{Perez}}, \bibnamefont{and}
  \bibinfo{author}{\bibfnamefont{D.}~\bibnamefont{Plouffe}},
  \bibinfo{journal}{\prb} \textbf{\bibinfo{volume}{66}},
  \bibinfo{pages}{075129} (\bibinfo{year}{2002}).

\bibitem[{\citenamefont{Ellzey et~al.}(2003)\citenamefont{Ellzey, Jr., and
  Villagran}}]{ellzey03}
\bibinfo{author}{\bibfnamefont{M.~L.} \bibnamefont{Ellzey}},
  \bibinfo{author}{\bibnamefont{Jr.}}, \bibnamefont{and}
  \bibinfo{author}{\bibfnamefont{D.}~\bibnamefont{Villagran}},
  \bibinfo{journal}{J. Chem. Inf. Comput. Sci.} \textbf{\bibinfo{volume}{43}},
  \bibinfo{pages}{1763} (\bibinfo{year}{2003}).

\bibitem[{\citenamefont{Yamamoto et~al.}(2005)\citenamefont{Yamamoto, Watanabe,
  and Watanabe}}]{yamamoto05}
\bibinfo{author}{\bibfnamefont{T.}~\bibnamefont{Yamamoto}},
  \bibinfo{author}{\bibfnamefont{K.}~\bibnamefont{Watanabe}}, \bibnamefont{and}
  \bibinfo{author}{\bibfnamefont{S.}~\bibnamefont{Watanabe}},
  \bibinfo{journal}{Phys. Rev. Lett.} \textbf{\bibinfo{volume}{95}},
  \bibinfo{pages}{065501} (\bibinfo{year}{2005}).

\bibitem[{\citenamefont{White et~al.}(1989)\citenamefont{White, Scalapino,
  Sugar, Loh, Gubernatis, and Scalettar}}]{white89}
\bibinfo{author}{\bibfnamefont{S.~R.} \bibnamefont{White}},
  \bibinfo{author}{\bibfnamefont{D.~J.} \bibnamefont{Scalapino}},
  \bibinfo{author}{\bibfnamefont{R.~L.} \bibnamefont{Sugar}},
  \bibinfo{author}{\bibfnamefont{E.~Y.} \bibnamefont{Loh}},
  \bibinfo{author}{\bibfnamefont{J.~E.} \bibnamefont{Gubernatis}},
  \bibnamefont{and} \bibinfo{author}{\bibfnamefont{R.~T.}
  \bibnamefont{Scalettar}}, \bibinfo{journal}{Phys. Rev. B}
  \textbf{\bibinfo{volume}{40}}, \bibinfo{pages}{506} (\bibinfo{year}{1989}).

\bibitem[{\citenamefont{Zhang and Callaway}(1989)}]{zhang89}
\bibinfo{author}{\bibfnamefont{Y.}~\bibnamefont{Zhang}} \bibnamefont{and}
  \bibinfo{author}{\bibfnamefont{J.}~\bibnamefont{Callaway}},
  \bibinfo{journal}{Phys. Rev. B} \textbf{\bibinfo{volume}{39}},
  \bibinfo{pages}{9397} (\bibinfo{year}{1989}).

\bibitem[{\citenamefont{Hirsch}(1983)}]{hirsch83}
\bibinfo{author}{\bibfnamefont{J.~E.} \bibnamefont{Hirsch}},
  \bibinfo{journal}{Phys. Rev. B} \textbf{\bibinfo{volume}{28}},
  \bibinfo{pages}{4059} (\bibinfo{year}{1983}).

\bibitem[{\citenamefont{Blankenbecler et~al.}(1981)\citenamefont{Blankenbecler,
  Scalapino, and Sugar}}]{blankenbecler81}
\bibinfo{author}{\bibfnamefont{R.}~\bibnamefont{Blankenbecler}},
  \bibinfo{author}{\bibfnamefont{D.~J.} \bibnamefont{Scalapino}},
  \bibnamefont{and} \bibinfo{author}{\bibfnamefont{R.~L.} \bibnamefont{Sugar}},
  \bibinfo{journal}{Phys. Rev. D} \textbf{\bibinfo{volume}{24}},
  \bibinfo{pages}{2278} (\bibinfo{year}{1981}).

\bibitem[{\citenamefont{Scalettar et~al.}(1993)\citenamefont{Scalettar,
  Dagotto, Bergomi, Jolicoeur, and Monien}}]{scalettar93}
\bibinfo{author}{\bibfnamefont{R.~T.} \bibnamefont{Scalettar}},
  \bibinfo{author}{\bibfnamefont{E.}~\bibnamefont{Dagotto}},
  \bibinfo{author}{\bibfnamefont{L.}~\bibnamefont{Bergomi}},
  \bibinfo{author}{\bibfnamefont{T.}~\bibnamefont{Jolicoeur}},
  \bibnamefont{and} \bibinfo{author}{\bibfnamefont{H.}~\bibnamefont{Monien}},
  \bibinfo{journal}{Phys. Rev. B} \textbf{\bibinfo{volume}{47}},
  \bibinfo{pages}{12316} (\bibinfo{year}{1993}).

\bibitem[{\citenamefont{Lammert and Rokhsar}(1993)}]{rokhsar93}
\bibinfo{author}{\bibfnamefont{P.~E.} \bibnamefont{Lammert}} \bibnamefont{and}
  \bibinfo{author}{\bibfnamefont{D.~S.} \bibnamefont{Rokhsar}},
  \bibinfo{journal}{Phys. Rev. B} \textbf{\bibinfo{volume}{48}},
  \bibinfo{pages}{4103} (\bibinfo{year}{1993}).

\end{thebibliography}

\end{document}